\documentclass[a4paper,11pt,onecolumn]{article}
\usepackage{amsmath}
\usepackage{amsfonts}
\usepackage{amssymb}
\usepackage{graphicx}
\usepackage{bm}
\usepackage{geometry}
\usepackage{cite}
\usepackage[makeroom]{cancel}
\usepackage[usenames,dvipsnames]{color}
\usepackage[plainpages=false,hypertexnames=true,hyperindex=true,pagebackref=false,raiselinks=true]{hyperref}

\input{tcilatex}
\graphicspath{{./Figures/}}

\begin{document}

\title{Gravitational quantum collapse in dilute systems}
\author{F. Lalo\"{e}\thanks{%
laloe@lkb.ens.fr} \\
Laboratoire Kastler Brossel, ENS-Universit\'e PSL,\\
CNRS, Sorbonne Universit\'e, Coll\`ege de France,\\
24 rue Lhomond 75005\ Paris, France}
\date{\today }
\maketitle

\begin{abstract}
Penrose  has suggested that large fluctuations of the  gravitational energy of quantum systems, resulting from fluctuations
of its density in space, may induce a quantum collapse mechanism \cite%
{Penrose-1996}, but he did not propose a precise dynamics for this process. We use the GBC (Gravitational Bohmian Collapse) model \cite{GBC}, which provides such a dynamics. The effects of collapse in dilute quantum systems are investigated, both in ordinary 3D space and in configuration space.
 We first discuss  how a single result appears during a quantum measurement. The GBC model predicts a continuous but very fast  evolution of the state vector that, at the end of the measurement, reproduces the von Neumann projection postulate. This ensures that the model remains compatible with the  relativistic nosignaling constraint. In the absence of any measurement, we study the spontaneous effects of the GBC process, which depend on the quantum correlation function of observables with the spatial density operator.
If the selected observable is the local current of the density fluid, we show that the collapse term leads to modifications of the Newton force, in a scalar or tensor form.
\end{abstract}

In quantum mechanics, the so called \textquotedblleft measurement problem\textquotedblright\ arises because the Schr\"{o}dinger equation alone is not able to predict the emergence of a single result in a quantum measurement experiment. Instead, and as noted in 1933 by von Neumann in his famous book  \cite{von-Neumann}, it predicts the appearance of a QSMDS (Quantum Superposition of Macroscopically Different States) containing all possible results at the same time. This superposition propagates further and further into the environment, without ever resolving into a single macroscopic component. The difficulty can be overcome  by various interpretations. The Copenhagen interpretation for instance limits the validity of the Schr\"{o}dinger equation by introducing  a \textquotedblleft cut\textquotedblright , or \textquotedblleft shifty split\textquotedblright , between the measured system and the macroscopic measurement apparatus, and then stating that the equation should not be used beyond this cut. Von Neumann proposed another solution by introducing his \textquotedblleft projection postulate\textquotedblright , which is nowadays introduced in most textbooks on quantum mechanics. Many different interpretations have been proposed by various authors. Still another solution is to modify the theory and  the Schr\"{o}dinger equation in order to resolve QSMDS.

In 1996, Penrose \cite{Penrose-1996} suggested that QSMDS could indeed be spontaneously
resolved into a single macroscopic component under the effect of
gravitational attraction. Introducing such a mechanism of  gravitational
quantum collapse into the theory would certainly solve the difficulties mentioned above; for instance the famous\ Schr\"{o}dinger cat paradox would immediately vanish.\
Penrose nevertheless proposed no specific mechanism for this collapse,
relying just on a qualitative argument involving the fluctutations of the gravitational energy of the system and the time-energy Heisenberg
uncertainty relation. A more precise mechanism was proposed in \cite{GBC},
with a simple model where the source of gravitational attraction is the
Bohmian positions of the particles, and where the gravitational constant
contains a small imaginary part to produce the collapse. As it is, the model remains
naive since, for instance, it uses the dBB (de Broglie-Bohm) theory to
obtain the motion of the particles, which means that it is non-relativistic.
We will call it the GBC (Gravitational Bohmian Collapse) model.

Modifying the standard Schr\"{o}dinger dynamics to include a collapse
mechanism is, of course, not a new idea. In 1986, Ghirardi, Rimini and Weber
introduced the GRW theory \cite{GRW}, where sudden random spatial
localization processes are added into the quantum dynamics; a stochastic
non-Hamiltonian term is introduced into the Schr\"{o}dinger equation.\ Soon after, a similar but continuous dynamics was
proposed in 1989 by Pearle with the CSL theory \cite{CSL,
Ghirardi-Pearle-Rimini}, based on the inclusion of Wiener processes in the
quantum dynamics.\ An interesting feature of these theories is that they
make predictions that differ from those of standard quantum mechanics,
which means that they can be tested experimentally \cite{Bassi-Lochan-et-al}.\ Another common feature is that they require the introduction of two
dimensional constants, usually a localization length and a time constant.

In 1989, Di\'{o}si proposed to relate dynamical collapse theories to gravity
in order to suppress the necessity for the introduction of these new
constants, and derive a collapse dynamics depending only of the Newton
constant.\ Nevertheless, Ghirardi, Grassi and Rimini soon showed \cite{GGR-1990} that this theory
makes predictions that are contradictory with known facts, in particular  in nuclear physics.\ The conclusion at the time was that, even if the Newton
gravitational constant is used in the theory, at least another dimensional
constant has to be introduced into the dynamics in order to avoid
contradictions with known experimental facts.

By contrast, the GBC\ gravitational collapse mechanism introduces no
dimensional parameter, but just one dimensionless constant $\varepsilon $%
: the Newton gravitational constant $G$ is replaced by $G^{\prime
}=G(1-i\varepsilon )$\ where $\varepsilon \simeq 10^{-3}$. Another
difference is that the dynamics of the collapse is a collective effect,
while in the GRW theory all particles in the physical system undergo
independent localization processes in parallel. In GBC dynamics, all Bohmian
positions of the particles contribute to the gravitational potential, which
in turn tends to localize each particle in the regions of low values of the potential.
Several consequences result from this collective character.\ First,  for
macroscopic systems, small
values of $\varepsilon $ are sufficient to obtain a very fast collapse.\ Second, the border between microscopic and macroscopic
physical systems is very sharp \cite{GBC}, since for a system of typical size $\ell$
the collapse time constant varies as $\ell^{5}$. Microscopic systems, for
which the gravitational self-attraction is extremely small, are almost not
perturbed by the collapse process; macroscopic systems undergo very fast
collapse. As a result, only situations involving a QSMDS in space (large
quantum fluctuation of the density  in ordinary space) are subject to rapid collapse.

Another feature of GBC is that the model does not involve any fluctuating
term in the dynamics, or Wiener process having an infinite spectrum.\ The GBC\ dynamics
is continuous and deterministic, the only random component being the
initial values of all Bohmian positions in the system. This is similar with the non-Markovian collapse model proposed by Tilloy and Wiseman \cite{Tilloy-Wiseman}, where the evolution of the conditional state vector of the system S is controlled by the evolution of the Bohmian positions of the particles in a bath entangled with S (but with no role of gravity). But this is also  completely different from the
other proposals relating gravity and quantum collapse,
which involve random functions: for instance Pearle and Squires use a theory
where the field causing collapse is the gravitational curvature scalar
containing a white noise fluctuating source \cite{Pearle-Squires}; Tilloy
and Di\'{o}si build a semiclassical gravity in the Newtonian limit where the
source of gravity contains a white noise component \cite{Tilloy-Diosi};
Adler proposes to add a complex part in the classical space-time metric,
which is somewhat similar to GBC, but this part is a fluctuating noise \cite%
{Adler-2016}; similar ideas have been developed by Gasbarri et al \cite%
{Gasbarri-et-al-2017}. Generally speaking, it is well-known that any
modification of the Schr\"{o}dinger dynamics should be combined with some
random element in the dynamics to avoid the possibility of faster than light
communication \cite{Gisin-1989,Bassi-Hejazi}.\ With GBC, this random
component is the initial Bohmian positions of the particles,
which ensures the required nonsignaling property (\S~\ref{superluminal}).

In this article, we study a few physical predictions of the GBC\
model.\ The reason why we focus on dilute systems is as follows. Within a piece of solid sitting in a well defined
region of space (no QSMDS), the gravitational constant created by all its
particles varies slowly inside its volume, with a maximum at the center of
the solid.\ In GRW and CSL theories, every particle inside the solid is
subject to localization processes that localize it within a microscopic
length (of the order of $10^{-7}$ m), which increases it energy; this
constantly heats up the solid, even if at a very small rate.\ Moreover the
bulk matter properties of the solid are changed \cite{Diosi-2014}.\ Within the
GBC\ dynamics, the localization length is much larger, since it is of the
order of the size $\ell $ of the macroscopic object, which means that the
energy transfer and the heating rate are much smaller.\ The net effect is rather a
collapse of the piece of solid inside itself, an effect that is
counterbalanced by the finite compressibility of the solid, so that the
density change remains extremely small.\ In a gas, or in a dilute
cloud of solid objects, the situation is different: no restoring force
opposes the density condensation induced by the collapse, and much larger
effects can be expected.

In \S~\ref{dynamics-model}, we briefly summarize the basic equations of the GBC dynamics.
In \S~\ref{measurements}, we come back to a subject that was already
studied in~\cite{GBC}, the effects of gravitational collapse during quantum
measurements, and the absence of superluminal signaling. In \S~\ref{dilute-systems}, we study spontaneous effects occuring in situations other
than quantum measurement, and calculate the evolution of the local density and current of the probability fluid in ordinary 3D space. In particular, we show how the localization process may induce a change of the gravitational constant. In \S~\ref{probability-fluid}, we study the motion of the probability fluid in configuration space, and discuss the validity of quantum equilibrium in various situations.

\section{Dynamics of the model}
\label{dynamics-model}

In this section, we briefly recall the main equations of the GBC dynamics.
The Hamiltonian $H$ of a physical system is the sum of its
usual Hamiltonian $H_{\text{int}}$ and of a gravitational Hamiltonian $%
H_{G}$, due to the attraction of masses with mass density $n_{G}(%
\mathbf{r})$:
\begin{equation}
H=H_{\text{int}}+H_{G}  \label{grav-1}
\end{equation}%
where $H_{G}$ is given by:%
\begin{equation}\label{grav-2}
H_{G}=-gGm \int \text{d}^{3}r~\Psi ^{\dagger }(\mathbf{r})\Psi (\mathbf{r}%
)\int \text{d}^{3}r^{\prime }~\frac{1}{\left\vert \mathbf{r}-\mathbf{r}%
^{\prime }\right\vert }n_{G}(\mathbf{r}^{\prime })
\end{equation}%
For the moment, we set $g=1$; $G$ is Newton's constant, $m$
the mass of the particles, and $\Psi (\mathbf{r})$ the quantum field
operator of the particles contained in the physical system. For the sake of simplicity, we have written the source of the gravitational field as the mass density $n_{G}(\mathbf{r}^{\prime })$ at the same time. An obvious improvement of (\ref{grav-2}) would be to insert a retarded value of the potential by replacing this mass density by $ n_{G}(\mathbf{r}^{\prime }, t - \left\vert \mathbf{r}-\mathbf{r}%
^{\prime }\right\vert /c)$, where $c$ is the speed of light. Since this does not change de discussion
in the rest of this article, the simple form (\ref{grav-2}) will suffice.

By setting  $n_{G}(\mathbf{r}^{\prime }) $ equal to the quantum local density average $< \Psi ^{\dagger }(\mathbf{r})\Psi (\mathbf{r}%
) >$, we would obtain the usual Schr\"{o}dinger-Newton equation \cite{Bahrami-Bassi-et-al}; but we will proceed differently.
We now introduce two non-standard
assumptions concerning the Hamiltonian $H_{G}$ describing the internal
gravitational attraction inside the system. First, we assume that $n_{G}(\mathbf{r})$ is
determined by the Bohmian positions $q_{n}$ of the $N$ particles of the system:%
\begin{equation}
n_{G}(\mathbf{r})=m\sum_{n=1}^{N}\delta (\mathbf{r}-\mathbf{q}_{n})  \label{grav-3}
\end{equation}%
Second, as in Ref. \cite{Diosi-Papp-2009}, we assume that the dimensionless constant $g$ contains a small
imaginary part $\varepsilon $:%
\begin{equation}
g=1-i\varepsilon   \label{grav-5}
\end{equation}%
The operator $H_{G}$ is now the sum of an Hermitian part $H_{G}^0 = H_{G} (\varepsilon =0 )$ and an antiHermitian part $iL$, where $L$ is the localization operator:
\begin{equation}\label{grav-6}
L=\varepsilon Gm\int \text{d}^{3}r~\Psi ^{\dagger }(\mathbf{r})\Psi (\mathbf{%
r})\int \text{d}^{3}r^{\prime }~\frac{1}{\left\vert \mathbf{r}-\mathbf{r}%
^{\prime }\right\vert }n_{G}(\mathbf{r}^{\prime })
\end{equation}%
This operator is diagonal in the position representation.


The non-normalized (if $\varepsilon \neq 0$) state vector $\left\vert \Phi (t)\right\rangle $ evolves according to:%
\begin{equation}
i\hbar \frac{\text{d}}{\text{d}t}\left\vert \Phi (t)\right\rangle =\left[ H_{%
\text{int}}+H_{G}^0 + iL \right] \left\vert \Phi (t)\right\rangle  \label{grav-7}
\end{equation}%
After normalization, this  state becomes a state $\left\vert \overline{\Phi }(t)\right\rangle$, which evolves according to \cite{GBC}:%
\begin{align}
i\hbar \frac{\text{d}}{\text{d}t}\left\vert \overline{\Phi }(t)\right\rangle
=\Bigg[ H_{\text{int}}&+H_{G}^{0}
\nonumber \\
 &+ i\varepsilon Gm \int \text{d}^{3}r\int
\text{d}^{3}r^{\prime }~\left[ \Psi ^{\dagger }(\mathbf{r})\Psi (\mathbf{r}%
)-D_{\Phi}(\mathbf{r})\right] ~\frac{1}{\left\vert \mathbf{r}-\mathbf{r}%
^{\prime }\right\vert }\;n_{G}(\mathbf{r}^{\prime })\Bigg] \left\vert \overline{\Phi }(t)\right\rangle  \label{grav-11}
\end{align}
where $D_{\Phi}(\mathbf{r})$ is the density of particles in state $\left\vert \overline{\Phi }(t)\right\rangle $:
\begin{equation}\label{defn-dphi}
D_{\Phi}(\mathbf{r})=
\left\langle \overline{\Phi }(t)\right\vert
\Psi ^{\dagger }(\mathbf{r})\Psi (\mathbf{r})
\left\vert \overline{\Phi }(t)\right\rangle
\end{equation}

%
%

\section{Fast gravitational collapse during measurements\label%
{measurements}}

Technically, the GBC model obtains a collapse of the state vector by enriching the standard
Schr\"{o}dinger dynamics with the addition of a point $P$ in configuration space; the
components of $P$ are all Bohmian positions of the particles contained in
the physical system. \ The role of $P$ is somewhat similar to that of the wave singularity in the de Broglie theory of the double solution \cite{De-Broglie-1927}. \ The position of $P$ is guided by the wave function, but conversely $P$ reacts on it (as opposed to what happens in usual dBB theory). One can then expect that this point should play the role of an attractor and, if the
number of particles is sufficient, that a collective effect might occur that forces
the $N$ body wave function to remain in the vicinity of $P$. We will see that, during a
measurement process, this creates the equivalent of the von Neumann
projection postulate, with a very sudden projection obtained within a
 continuous dynamics.

For the sake of simplicity, we assume that the particles are spinless;
adding spins within the Pauli spin theory would not change much the
discussion, while making the notation more complicated.

\subsection{Appearance of a single result during a quantum measurement}\label{single-result}

In the initial stage of a quantum measurement experiment, the measured system $S$ becomes
entangled with the measurement apparatus $M$. The usual de
Broglie-Bohm (dBB) theory then predicts that the Bohmian positions of the
particles of $M$ begin to play an important role in the dynamics \cite{Tastevin-Laloe}.\ These positions are for instance the positions of the particles inside the pointer of the
measurement apparatus. Together with the positions attached to the
measured quantum system $S$, they determine the position of  $P$.

Within standard dBB theory, during the interaction
between $S$ and $M$, the
point $P$ can a priori follow several branches of the wave function (regions of the configuration space where it does not vanish). Each branch is
associated with a single measurement result.\ Because of the cohesive forces between the
particles inside the pointer, all
individual Bohmian positions have to remain grouped together
 in one branch of the state vector. This is because the position of the point $P$ in configuration space cannot reach points where the many-body wave function vanishes, in particular points where the positions variables of the particles inside the pointer are spread among different regions of space.
 Indeed, within standard dynamics,
the pointer may end up in a superposition of states indicating different
results, but cannot
spontaneously decompose into a broken pointer state, a superposition of pointer fragments located at different places.
 Moreover, as soon as the overlap between the wave functions of these branches tends to zero (this happens because the particles inside the pointer move in different directions in  different branches), all the Bohmian positions have to remain inside the same branch until the end of the experiment. Finally, if several realizations of the experiment are performed, in successive realizations  the positions may follow different branches;
 the standard dBB dynamics predicts that the proportion of cases where all positions follow a given branch is given by the Born rule.

Within the GBC dynamics, what happens during a single realiztion of an experiment? Before the measurement process starts, the dynamics of an isolated quantum systems $S$  remains in practice indistinguishable from the standard quantum dynamics in the absence of gravity. It is well known that the gravitational interaction between two protons is about $10^{38}$  times smaller than the electromagnetic interaction; in addition, the localization term is multiplied by a small parameter  $\varepsilon \simeq 10^{-3}$.
During the early stages of a measurement,  and as long as the entanglement between the measured system and its environment (including the measurement apparatus $M$) remains microscopic, the effect of the localization term still remains completely negligible.
But, within $M$, successive interactions between neighbour  particles make the entanglement progress rapidly, over a distance $\ell (t)$. This distance can be assumed to increase roughly linearly in time:
$\ell (t) \simeq v t$, where $v$ is for instance the velocity of sound inside a solid.
Now, the inverse time constant $1 / \tau $ for collapse increases as the fifth power of $\ell (t)$ \cite{GBC}, so that $1 / \tau \sim (vt)^5$.
 This means that the rate of cancellation of an empty  branch of the QSMDS (a branch that does not contain Bohmian positions) increases  very rapidly, as the fifth power of time.
 At a certain point, the progression of entanglement in the measurement apparatus crosses the point where, by a collective gravitational effect inside $M$, the collapse dynamics becomes very fast and  selects the only non-empty component;  the transition from one regime to the other is almost instantaneous.
In the limit where the process is really instantaneous, we obtain a result that is perfectly equivalent to the  standard von Neumann projection postulate.

\subsection{No superluminal communication}\label{superluminal}

As recalled in the introduction, changing the Schr\"{o}%
dinger dynamics, for instance by adding a nonlinear term in the equation,
may introduce the possibility of superluminal signaling \cite{Gisin-1989,Bassi-Hejazi}, in
contradiction with relativity.\ A first method
to avoid this problem is to introduce stochasticity into the dynamics: in
this way, the perturbation introduced by the nonlinear term becomes random,
and one can show that it cannot carry messages.\ This is for example the case in GRW and CSL
theories.\ The GBC\ model makes use of a second possibility: the dynamics
remains deterministic, but the random component is introduced by the
initial Bohmian positions of all particles, those of $S$ as well as those of $M$.

Assume that Alice and Bob do experiments in two remote galaxies, and share an entangled pair of particles. As long as neither of them makes her/his particle interact with a macroscopic system, for instance by making a measurement, no difference with standard quantum mechanics can be observed. This is because, as we have seen,  for microscopic systems the GBC localization term remains completely negligible when compared to the standard Hamiltonian.  In any case,  at this stage of the experiment, the nosignaling relativistic condition does not yet really apply: it forbids instantaneous communication at a macroscopic level between macroscopic observers, but not possible influences   at a microscopic level.

The situation changes completely when Alice starts a measurement. Her particle then becomes entangled with another quantum system $M$ that is macroscopic, and a QSMDS is initiated. We have seen above that the rapid propagation of entanglement within $M$, by successive interactions between neighbour  particles, introduces a resolution rate of the QSMDS that varies as the fifth power of time. This results in a projection dynamics that is very sudden. Moreover, it is governed by the initial positions of the Bohmian particles in Alice's laboratory, which she cannot control or determine (and which Bob cannot know either), and which reproduce the Born rule statistically. At the end, the evolution of the state vector is in practice
equivalent to that predicted by the instantaneous  von Neumann
projection postulate. Since this postulate is known to be free of the superluminal communication problem, the GBC model does not have this problem either.

Another way to reach the same conclusion is to analyse how Alice, by choosing the kind of measurement she will perform, can influence  the collapse process of the common state vector she shares with Bob, and send him a message in this way. But neither she, nor Bob, can know or control the initial value of the Bohmian positions in her lab. So, the information Alice can send to Bob by projecting the state vector  has to be averaged over these positions. But this average is mathematically equivalent to the usual partial trace operation that determines,  in standard theory, the density operator received by Bob. It is known that this leads to an operator that is independent of what Alice does, which means that no information is transmitted instantaneously.

\subsection{Quantum equilibrium}
\label{quantum-equilibrium}

Standard dBB theory predicts that the position distribution of the
particles, averaged over many realizations of an experiment, obeys the so
called \textquotedblleft quantum equilibrium\textquotedblright\ property: it
remains constantly equal to the quantum distribution of probability given by
the square of the wave function (in configuration space).\ Valentini has
shown \cite{Valentini-1991,Valentini-2002} that any deviation from quantum equilibrium would
introduce a possibility of superluminal communication:   this
equilibrium is a necessary condition for avoiding a contradiction with
relativity.\ Moreover, Valentini et al.
have shown \cite{Valentini-Westman-2004,Towler-Russel-Valentini} that,
within the dBB dynamics, the quantum equilibrium is an attractor of the dBB
dynamics: if, for some reason, the distribution of the positions deviates
from the quantum distribution, a fast quantum relaxation process constantly
restores quantum equilibrium.

Since GBC changes the evolution of the wave function, and therefore the
quantum distribution, quantum equilibrium is no longer necessarily
obeyed at any time. In a \textquotedblleft normal\textquotedblright\
situation (no QSMDS), the deviation from quantum equilibrium remains
completely negligible: it is created by a gravitational attraction term,
which is usually completely ignored in quantum mechanical calculations, and moreover
multiplied by a small coefficient $\varepsilon$. In addition, it is constantly counterbalanced
by the quantum relaxation process, which tends to restore the equilibrium.\

But now consider a situation where standard quantum theory predicts the
appearance of a QSMDS.\ As we have seen above, this is the case, for instance, in the initial
stage of a quantum measurement experiment, when the measured physical system
$S$ becomes entangled with the various components of the measurement
apparatus $M$.  The wave function $\Phi $ is strongly
attracted towards the region of space where all Bohmian positions move
together. This of course completely changes the quantum distribution $%
\left\vert \Phi \right\vert ^{2}$ in  configuration space, which
suddenly destroys the quantum equilibrium between the two distributions.
But this disappearance of the quantum equilibrium remains a short transient
effect: immediately after the collapse of  $\Phi $ has taken place, the QSMDS has disappeared, and the quantum
relaxation process discussed by Valentini et al.\ \cite%
{Valentini-Westman-2004,Towler-Russel-Valentini} tends to restore the
equilibrium. A situation similar to the one  before measurement is then quickly
recovered.

To summarize, strong deviations from quantum equilibrium may happen
during a short time, when the two components of $\Phi $ located in two
different regions of space begin to appear.  But, as soon as the spatial separation becomes large, the equilibrium is promptly
restored. In \S~\ref{probability-fluid}, we come back in
more detail to this phenomenon in configuration space.

\section{Slow localization effects in dilute systems}
\label{dilute-systems}

The effects of the Hermitian part of the Bohmian gravitational Hamiltonian have already been discussed in several articles, see for instance \cite{Struyve, Struyve-2,Peter-Pinho-et-al,Pinho-Pinto} and references contained. Here we focus on the physical effects of the antiHermitian part in situations other than quantum measurements, already discussed in \S~\ref{single-result}.
 Instead of liquid or  solid systems, we
consider dilute physical systems having lower densities, so that the
effects of the gravitational collapse are much weaker. Moreover, in dense systems, forces between the
particles constantly and efficiently tend to oppose changes of the density, which does not happen in dilute systems such as gases. As a result we can expect that, instead of producing a very fast collapse as during a quantum measurement, the gravitational  localization term in a dilute system will produce softer and more continuous effects, occurring over much longer time scales.

\subsection{Localization terms}

 The Bohmian positions $q_{n}$ of the $N$ particles of the system are
the sources of the gravitational potential $\mathcal{V}_{G}(\mathbf{r})$:%
\begin{equation}
\mathcal{V}_{G}(\mathbf{r})=-Gm^{2}\sum_{n=1}^{N}\frac{1}{\left\vert \mathbf{%
r}-\mathbf{q}_{n}\right\vert }  \label{GBC-1}
\end{equation}%
 Equation (\ref{grav-11}) shows that, in addition to the standard Hamiltonian term (including the gravitational potential $\mathcal{V}_{G}(\mathbf{r})$),
the evolution of the normalized state vector $\left\vert \overline{\Phi }(t)\right\rangle $ contains a localization term that reads:%
\begin{equation}
\left. \frac{\text{d}}{\text{d}t}\right\vert _{\text{loc}}\left\vert \overline{\Phi }(t) \right\rangle =\frac{\varepsilon }{\hbar }\int \text{d}^{3}r^{\prime }~%
\left[ D(\mathbf{r}^{\prime })-\left\langle D(\mathbf{r}^{\prime
})\right\rangle \right] ~\;\left\vert \mathcal{V}_{G}(\mathbf{r}^{\prime
})\right\vert ~\left\vert \overline{\Phi }(t)\right\rangle  \label{GBC-2}
\end{equation}%
where  $D(\mathbf{r})$ is the operator associated with the local density of particles:
\begin{equation}
D(\mathbf{r})=\Psi ^{\dagger }(\mathbf{r})\Psi (\mathbf{r})  \label{GBC-3}
\end{equation}
%
and
 $\left\langle D(\mathbf{r}%
)\right\rangle $ its average value in state $\left\vert \overline{\Phi }%
(t)\right\rangle $.
Relation (\ref{GBC-2}) indicates that the spontaneous localization
process is more effective in regions of space where $\left\vert \mathcal{%
V}_{G}(\mathbf{r})\right\vert $ takes large values. It tends to increase
the value of the wave function if the action of the quantum density operator
$D(\mathbf{r})$ exceeds that of the multiplication by the local average of
this density, to decrease it otherwise.

The evolution of $\left\langle D(\mathbf{r})\right\rangle $ due to the
localization process is then given by the relation:%
\begin{equation}
\left. \frac{\text{d}}{\text{d}t}\right\vert _{\text{loc}}\left\langle D(%
\mathbf{r})\right\rangle =\frac{2\varepsilon }{\hbar }\int \text{d}%
^{3}r^{\prime }~\left[ \left\langle D(\mathbf{r})D(\mathbf{r}^{\prime
})\right\rangle -\left\langle D(\mathbf{r})\right\rangle \left\langle D(%
\mathbf{r}^{\prime })\right\rangle \right] ~\left\vert \mathcal{V}_{G}(%
\mathbf{r}^{\prime })\right\vert  \label{GBC-5}
\end{equation}%
Since:%
\begin{equation}
\int \text{d}^{3}r~D(\mathbf{r})=N  \label{GBC-6}
\end{equation}%
it is easy to check that the localization term does not change the total
number of particles, as expected.\ Equation (\ref{GBC-5}) shows that all
points of space $\mathbf{r}^{\prime }$ contribute to the variation ot the
number density at any point $\mathbf{r}$, with a weight proportional to the
gravitational potential at point $\mathbf{r}^{\prime }$, but also with a
coefficient that is proportional to the density correlation function at the
two points of space; the average over all space of this correlation
coefficient vanishes.


\subsection{Evolution of the density\label{density-evolution}}

We introduce the density correlation function $\left\langle F(\mathbf{r},\mathbf{r}^{\prime })\right\rangle $ by:%
\begin{equation}
\left\langle D(\mathbf{r})D(\mathbf{r}^{\prime })\right\rangle =\left\langle
D(\mathbf{r})\right\rangle \left\langle D(\mathbf{r}^{\prime })\right\rangle
+\left\langle F(\mathbf{r},\mathbf{r}^{\prime })\right\rangle  \label{GBC-9}
\end{equation}%
If we sum this relation over d$^{3}r$ and use relation (\ref{GBC-6}), we
obtain:%
\begin{equation}
 \int \text{d}^{3}r~\left\langle F(\mathbf{r},\mathbf{r}%
^{\prime })\right\rangle
 =
 \int \text{d}^{3}r^{\prime }~\left\langle F(\mathbf{r},\mathbf{r}^{\prime
})\right\rangle
  =0  \label{GBC-10}
\end{equation}%
This shows that the function $\left\langle F(\mathbf{r},\mathbf{r}^{\prime })\right\rangle$
necessarily takes positive and negative values. If $\mathbf{r}^{\prime }=%
\mathbf{r}$, it is positive since:%
\begin{equation}
\left\langle \left[ D(\mathbf{r})\right] ^{2}\right\rangle >\left\langle D(%
\mathbf{r})\right\rangle ^{2}  \label{GBC-11}
\end{equation}%
so that:%
\begin{equation}
\left\langle F(\mathbf{r},\mathbf{r})\right\rangle >0  \label{GBC-12}
\end{equation}%
When $\left\vert \mathbf{r}^{\prime }-\mathbf{r}\right\vert \rightarrow
\infty $, the quantum correlations disappear, and: \
\begin{equation}
\left\langle F(\mathbf{r},\mathbf{r}^{\prime })\right\rangle \underset{%
\left\vert \mathbf{r}^{\prime }-\mathbf{r}\right\vert \rightarrow \infty }{%
\rightarrow }0  \label{GBC-13}
\end{equation}%

Equation (\ref{GBC-5}) reads:%
\begin{equation}
\left. \frac{\text{d}}{\text{d}t}\right\vert _{\text{loc}}\left\langle D(%
\mathbf{r})\right\rangle =\frac{2\varepsilon }{\hbar }\int \text{d}%
^{3}r^{\prime }\left\vert \mathcal{V}_{G}(\mathbf{r}^{\prime })\right\vert
~\left\langle F(\mathbf{r},\mathbf{r}^{\prime })\right\rangle  \label{GBC-14}
\end{equation}%
$\left\langle F(\mathbf{r},\mathbf{r}^{\prime })\right\rangle $ has a
positive peak at $\mathbf{r}^{\prime }=\mathbf{r}$, then negative values
when $\left\vert \mathbf{r}^{\prime }-\mathbf{r}\right\vert \simeq \lambda _{c}$
(where $\lambda _{c}$ is a correlation length), and then tends to zero
when $\left\vert \mathbf{r}^{\prime }-\mathbf{r}\right\vert \gg \lambda _{c}$.
 If we integrate this relation over d$^{3}r$
and use (\ref{GBC-10}), we recover the conservation law of the total number
of particles.\ The same relation shows that, if the gravitational potential $%
\mathcal{V}_{G}(\mathbf{r}^{\prime })$ is constant over space, no evolution
of the local density takes place. This evolution is therefore driven by the
gravitational gradient, i.e. the gravitational force, rather than the potential itself.

\subsection{Short correlation range}

To have an idea of the origin of  the correlation length $\lambda _c$, let us for instance assume that the system is described by a  number state containing many individual states with populations $n_1$, $n_2$, .., $n_p$:
\begin{equation}
\left\vert \overline\Phi _{0}\right\rangle =\frac{1}{\sqrt{n_{1}!n_{2}!...n_{p}!}}%
\prod_{k}(a_{1}^{\dagger })^{n_{1}}(a_{2}^{\dagger
})^{n_{2}}...(a_{p}^{\dagger })^{n_{p}}\left\vert \text{vac.}\right\rangle
\label{19}
\end{equation}%
where $\left\vert \text{vac.}\right\rangle$ is the vacuum state.
The density-density correlation operator is:%
\begin{equation}\label{21-transfere}
\widehat{D}(\mathbf{r})\widehat{D}(\mathbf{r'})
=
\sum_{k,l,k^{\prime },l^{\prime }}\varphi _{k}^{\ast }(\mathbf{r})\varphi
_{l}^{{}}(\mathbf{r})\varphi _{k^{\prime }}^{\ast }(\mathbf{r}^{\prime
})\varphi _{l^{\prime }}^{{}}(\mathbf{r}^{\prime })~a_{k}^{\dagger
}a_{l}~a_{k^{\prime }}^{\dagger }a_{l^{\prime }}
\end{equation}%
where $\varphi _k ( \mathbf r )$ is the wave function of the individual state created by $( a_k )^\dagger$. This leads to:
\begin{equation}\label{22-transfere}
\left\langle F(\mathbf{r},\mathbf{r'})\right\rangle=
\sum_{k\neq k^{\prime }}n_{k}\left(
n_{k^{\prime }}+1\right) \varphi _{k}^{\ast }(\mathbf{r})\varphi _{k^{\prime
}}^{{}}(\mathbf{r})\varphi _{k^{\prime }}^{\ast }(\mathbf{r}^{\prime
})\varphi _{k}^{{}}(\mathbf{r}^{\prime })
\end{equation}%
If $\mathbf{r}-%
\mathbf{r}^{\prime }$ becomes large, the sum of the products $\varphi
_{k}^{\ast }(\mathbf{r}^{\prime })\varphi _{k}^{{}}(\mathbf{r})$ tends to
zero by destructive interference between all the oscillating wave functions $\varphi _k (\mathbf{r})$.
The characteristic length of this cancellation at large distances is determined by the dispersion of the momenta of the various wave functions $\varphi _{k}^{{}}(\mathbf{r})$. Its  value depends on the physical system studied, but in many cases we can assume that $\lambda _c$ is a microscopic length.

We then assume that $\lambda _{c}$ is smaller than the charactetistic distances
over which $\mathcal{V}_{G}(\mathbf{r}^{\prime })$ varies; in particular, we
assume the absence of QSMDS such as those appearing for a short time
during a quantum measurement process, as discussed in \S\ \ref{measurements}%
.\ We can then expand the spatial variations of the potential and obtain:%
\begin{equation}\label{GBC-23}
\left. \frac{\text{d}}{\text{d}t}\right\vert _{\text{loc}}\left\langle D(%
\mathbf{r})\right\rangle =\frac{2\varepsilon }{\hbar }\bm\nabla \left\vert
\mathcal{V}_{G}(\mathbf{r})\right\vert \cdot \int \text{d}^{3}r^{\prime
}~\left( \mathbf{r}^{\prime }-\mathbf{r}\right) ~\left\langle F(\mathbf{r},\mathbf{r}^{\prime })\right\rangle  \end{equation}%
where $\bm\nabla \left\vert \mathcal{V}_{G}(\mathbf{r})\right\vert $ is
nothing but the local gravitational force $\mathbf{F}_{G}(\mathbf{r})$.

What appears in the integral \ in the right side of (\ref{GBC-23}) is only
the component $z^{\prime \prime }$ of $\mathbf{r}^{\prime }-\mathbf{r}$
along the direction of the gravitational force.\ If the function $F(\mathbf{r%
},\mathbf{r}+\mathbf{r}^{\prime \prime })$ is symmetric (even) with respect
to the plane $z^{\prime \prime }=0$, no evolution of the local density takes
place under the effect of spontaneous localization; it is therefore the asymmetry of
this function with respect to the plane $z^{\prime \prime }=0$ that drives
the evolution of $\left\langle D(\mathbf{r})\right\rangle $.

The order of magnitude of the maximum value of the right hand side   of relation (\ref{GBC-23}) is then:
\begin{equation}
\frac{\varepsilon }{\hbar  } \, \mathbf{F}_{G}(\mathbf{r})  \left\langle D(\mathbf{r})\right\rangle
 \int_{ \vert \mathbf{r} \vert \leq \lambda _c } \text d ^3 \mathbf{r}' \, \lambda _c
 \left\langle D(\mathbf{r}')\right\rangle
 \simeq
 \frac{\varepsilon  }{\hbar } \,\mathbf{F}_{G}(\mathbf{r}) \, \lambda _c ^4
 \left\langle D(\mathbf{r})\right\rangle ^2
 \end{equation}
The time constant of the relative variation of the local density is obtained by dividing this result by $\left\langle D(\mathbf{r})\right\rangle$. To obtain an order of magnitude, we may assume that the system is a gas in standard conditions with a number density  of $10 ^{26} \text m ^{-3}$, moving in the Earth gravitational field with $m=10^{-26}$kg; we take $\varepsilon = 10 ^{-3}$ and choose the microscopic length $\lambda _c = 10 ^{-9}$ m. We then get a large time constant of the order of $10 ^{6}$s, probably much too long to be experimentally detected. But this time constant varies very rapidly as a function of $\lambda _c$: if we assume that $\lambda _c \simeq 10 ^{-7}$m, we obtain a time constant of $10 ^{-2}$s, and of course even much shorter time constants if $\lambda _c$ is larger. The conclusion is that the results are extremely sensitive to the length over which the quantum fluctuations of the density extend. It is therefore difficult to make predictions without a more precise quantum model of the physical system.

\subsection{Evolution of the local current of particles\label{current}}

The operator associated with the local current of particles is:%
\begin{equation}
\mathbf{J}(\mathbf{r})=\frac{\hbar }{2mi}\left[ \Psi ^{\dagger }(\mathbf{r})%
\bm\nabla \Psi (\mathbf{r})-\bm\nabla \Psi ^{\dagger }(\mathbf{r})~\Psi (%
\mathbf{r})\right]  \label{GBC-18}
\end{equation}%
The evolution of its average value due to the localization term is:%
\begin{equation}
\left. \frac{\text{d}}{\text{d}t}\right\vert _{\text{loc}}\left\langle
\mathbf{J}(\mathbf{r})\right\rangle =\frac{\varepsilon }{\hbar }\int \text{d}%
^{3}r^{\prime }~\left[ \left\langle \mathbf{J}(\mathbf{r})D(\mathbf{r}%
^{\prime })+D(\mathbf{r}^{\prime })\mathbf{J}(\mathbf{r})\right\rangle
-2\left\langle \mathbf{J}(\mathbf{r})\right\rangle \left\langle D(\mathbf{r}%
^{\prime })\right\rangle \right] ~\left\vert \mathcal{V}_{G}(\mathbf{r}%
^{\prime })\right\vert  \label{GBC-19}
\end{equation}%
Now, what determines the evolution is the quantum correlation function
between the local density operator $D(\mathbf{r}^{\prime })$ and the local
current $\mathbf{J}(\mathbf{r})$. We then set:%
\begin{equation}
\left\langle \bm{J}(\bm{r})D(\bm{r}^{\prime })+D(\bm{r}^{\prime })\bm{J}(%
\bm{r})\right\rangle =2\left[ \left\langle \mathbf{J}(\mathbf{{r})}%
\right\rangle \left\langle D(\mathbf{r}^{\prime }\mathbf{)}\right\rangle
+\left\langle \mathbf{K}(\bm{r},\bm{r}^{\prime })\right\rangle \right]
\label{GBC-20}
\end{equation}%
where $\left\langle \mathbf{K}(\bm{r},\bm{r}^{\prime })\right\rangle $  depends on the quantum correlations between $\left\langle
\mathbf{J}(\mathbf{{r})}\right\rangle $ at the current point of space $%
\mathbf{r}$ and the value of the local density $D(\mathbf{r}^{\prime }%
\mathbf{)}$ at all other points of space.\ We then obtain:%
\begin{equation}
\left. \frac{\text{d}}{\text{d}t}\right\vert _{\text{loc}}\left\langle
\mathbf{J}(\mathbf{r})\right\rangle =\frac{\varepsilon }{\hbar }\int \text{d}%
^{3}r^{\prime }\left\vert \mathcal{V}_{G}(\mathbf{r}^{\prime })\right\vert
~\left\langle \mathbf{K}(\bm{r},\bm{r}^{\prime })\right\rangle
\label{GBG-21}
\end{equation}

An integration of (\ref{GBC-20}) over d$^{3}r^{\prime }$  provides, since the
integral of $D(\mathbf{r}^{\prime }\mathbf{)}$ over d$^{3}r^{\prime }$ is
equal to a constant $N$:
\begin{equation}
\int \text{d}^{3}r^{\prime }~\left\langle \mathbf{K}(\bm{r},\bm{r}^{\prime
})\right\rangle =0  \label{GBC-22}
\end{equation}%
which is similar to (\ref{GBC-10}).\ This shows that the components of $%
\mathbf{K}(\bm{r},\bm{r}^{\prime })$ take positive and negative values when $%
\mathbf{r}^{\prime }$ varies.\ Nevertheless, we do not have the equivalent
of the positivity relation (\ref{GBC-12}) with the local current. In any
case, relation (\ref{GBC-22}) ensures that the right hand side of (\ref%
{GBG-21}) vanishes if the gravitational potential $\mathcal{V}_{G}(\mathbf{r}%
^{\prime })$ is independent of $\mathbf{r}^{\prime }$.\ Therefore, if $%
\left\langle \mathbf{K}(\bm{r},\bm{r}^{\prime })\right\rangle $ has a
limited range, we can locally expand the variation of the gravitational
potential and write:%
\begin{equation}
\left. \frac{\text{d}}{\text{d}t}\right\vert _{\text{loc}}\left\langle
\mathbf{J}(\mathbf{r})\right\rangle =\frac{\varepsilon }{\hbar }\int \text{d}%
^{3}r^{\prime }~\left[ \mathbf{F}_{G}(\mathbf{r}) \cdot \left( \mathbf{r}^{\prime }-\mathbf{r}\right) \right]
\left\langle \mathbf{K}(\bm{r},\bm{r}^{\prime })\right\rangle  \label{GBC-24}
\end{equation}%
In this expression, the gravitational force $ \mathbf{F}_{G}(\mathbf{r})=   \bm\nabla \left\vert \mathcal{V}_{G}(%
\mathbf{r})\right\vert $ can be moved outside of
the integral.\ This shows that the localization process creates an
additional term in the time evolution of $\left\langle \mathbf{J}(\mathbf{r}%
)\right\rangle $ that is proportional to the usual gravitational force.

The\ gravitational force is therefore modified by the localization process,
but the additional term is not necessarily collinear with $\mathbf{F}_{G}(%
\mathbf{r})$: the relation is tensorial in general, and depends on the
properties of the correlation function $\left\langle \mathbf{K}(\bm{r},\bm{r}%
^{\prime })\right\rangle $. Nevertheless, if the correlation function is invariant by
rotation around $\mathbf{F}_{G}(\mathbf{r})$, the additional force is collinear to the standard gravitational force; the localization term then
just appears as a correction of the value of the Newton constant.

As above, evaluating the effect of the localization term quantitatively is difficult, because of the very fast dependence of the time constants on the correlation length $\lambda _c$. Nevertheless, effects on the local current seem more accessible to observation than direct effects on the density. In astrophysical systems for instance, even a small change of the velocity may, after  propagation during a very long time, lead to larger changes of the density.

\section{Probability fluid in configuration space\label{probability-fluid}}

We now consider the evolution of the probability fluid in the configuration
space.\ For simplicity, we consider an ensemble of spinless particles
described by the wave function $\Phi (\mathbf{r}_{1},\mathbf{r}_{2},..,%
\mathbf{r}_{N})$. In configuration
space, the density of the probability fluid  $\rho _{N}(\mathbf{r}_{1},\mathbf{r}_{2},..,\mathbf{r}_{N})$ is given
by:%
\begin{equation}
\rho _{N}(\mathbf{r}_{1},..,\mathbf{r}_{N})=\left\vert \Phi (\mathbf{r}%
_{1},..,\mathbf{r}_{N})\right\vert ^{2}  \label{GBC-25}
\end{equation}%
and the standard expression of the probability current $\mathbf{J}_{N}(%
\mathbf{r})$ in this space is:%
\begin{equation}
\mathbf{J}_{N}(\mathbf{r}_{1},..,\mathbf{r}_{N})=\frac{\hbar }{2mi}\left[
\Phi ^{\ast }(\mathbf{r}_{1},..,\mathbf{r}_{N})\bm\nabla _{N}\Phi (\mathbf{r}%
_{1},..,\mathbf{r}_{N})-\text{c.c.}\right]  \label{GBC-26}
\end{equation}%
where c.c. means \textquotedblleft complex conjugate\textquotedblright , and $\bm\nabla _{N}$ denotes the $N$ dimensional gradient. We now proceed to
calculate the value of the expression:%
\begin{equation}
\frac{\partial }{\partial t}\rho _{N}(\mathbf{r}_{1},..,\mathbf{r}_{N})+\bm%
\nabla _{N}\cdot \mathbf{J}_{N}(\mathbf{r}_{1},\mathbf{r}_{2},..,\mathbf{r}%
_{N})  \label{GBC-27}
\end{equation}%
In standard theory, the value of this expression is zero, which
expresses the local conservation of probability. In the gravitational
collapse model, since  the Schr\"{o}dinger equation is modified, the expression
no longer vanishes in general.

Within this model, the wave function $\Phi $ evolves according to:%
\begin{align}
\frac{\partial }{\partial t}\Phi (\mathbf{r}_{1},..,\mathbf{r}_{N})& =i\left[
\frac{\hbar}{2m} \Delta _{N}-V(\mathbf{r}_{1},..,\mathbf{r}_{N})\right] \Phi (\mathbf{r}%
_{1},..,\mathbf{r}_{N})  \notag \\
+\varepsilon & \left[ \sum_{n=1}^{N}\left\vert \mathcal{V}_{G}(\mathbf{r}%
_{n})\right\vert -\int \text{d}^{3}r^{\prime }~\left\langle D(\mathbf{r}%
^{\prime })\right\rangle \left\vert \mathcal{V}_{G}(\mathbf{r}^{\prime
})\right\vert \right] \Phi (\mathbf{r}_{1},..,\mathbf{r}_{N})  \label{GBC-28}
\end{align}%
where $\Delta _{N}$ denotes the $N$ dimensional Laplacian (the sum of $N$
individual Laplacians with respect to $\mathbf{r}_{n}$). This equation is
explicitly nonlinear since $\left\langle D(\mathbf{r}^{\prime })\right\rangle$ depends on $\Phi $:%
\begin{equation}
\left\langle D(\mathbf{r}^{\prime })\right\rangle = \sum_{n=1}^{N}\int
\text{d}^{3}r_{1}..\int \text{d}^{3}r_{n-1}..\int \text{d}^{3}r_{n+1}..\int \text{d}%
^{3}r_{N}~\left\vert \Phi (\mathbf{r}_{1},.., \mathbf{r}_{n}=\mathbf{r}^{\prime }  ,..,\mathbf{r}_{N})\right\vert
^{2} \label{GBC-29}
\end{equation}
Relation (\ref{GBC-26}) provides:%
\begin{equation}
\bm\nabla _{N}\cdot \mathbf{J}_{N}=\frac{\hbar }{2mi}\left[ \Phi ^{\ast }\bm%
\Delta _{N}\Phi -\text{c.c.}\right]  \label{GBC-30}
\end{equation}%
If we replace $\bm\Delta _{N}\Phi $ in this expression by the one resulting
from (\ref{GBC-28}), we obtain:%
\begin{equation}
\bm\nabla _{N}\cdot \mathbf{J}_{N}=\left\{ -\frac{\partial }{\partial t}%
+2\varepsilon \left[ \sum_{n=1}^{N}\left\vert \mathcal{V}_{G}(\mathbf{r}%
_{n})\right\vert -\int \text{d}^{3}r^{\prime }~\left\langle D(\mathbf{r}%
^{\prime })\right\rangle \left\vert \mathcal{V}_{G}(\mathbf{r}^{\prime
})\right\vert \right] \right\} \rho _{N}(\mathbf{r}_{1},..,\mathbf{r}_{N})
\label{GBC-31}
\end{equation}%

Consider now a distribution of Bohmian positions in configuration space
(distribution over many realizations of the experiment) that coincides with
the quantum distribution $\rho _{N}$. If $\varepsilon $ is equal to zero, the two distributions coincide again
at time $t+$d$t$; this is no longer true if $\varepsilon $ does not vanish. The term in $\varepsilon $ contains the difference between two different samplings of
the gravitational potential: the sum of all $\left\vert \mathcal{V}_{G}(%
\mathbf{r}_{n})\right\vert $ at the various 3D components $\mathbf{r}_{n}$ of the current point $P$
in configureation space, and the 3D average of $\left\vert
\mathcal{V}_{G}(\mathbf{r}^{\prime })\right\vert $ over the density $%
\left\langle D(\mathbf{r}^{\prime })\right\rangle $ associated with the
state vector at time $t$. Two different regimes are therefore possible.

\subsection{Quantum equilibrium in ordinary situations}

In ordinary situations (no QSMDS) one can expect that, for a macroscopic
 physical system, the very large value of $N$ ensures that, at least for
 typical configurations, the two samplings provide the same value (within
a fluctuation proportional to $\sqrt{N}$). So, in this case, two reasons contribute to make the effect of the term
in $\varepsilon $ completely negligible:

- the sampling of the gravitational potential in the two terms is almost the
same.

- the remaining component is proportional to the gravitational interaction.\
This interaction is much smaller than the terms in the standard Hamiltonian,
which constantly tend to restore  the equality
between the distribution of Bohmian positions and $\rho _{N}$ \cite%
{Valentini-Westman-2004,Towler-Russel-Valentini}.

Therefore, in ordinary situations, departure from
quantum equilibrium remains completely negligible. Quantum equilibrium is still
an excellent approximation.

\subsection{Transient departures from quantum equilibrium in QSMDS}

If a QSMDS occurs, and if the various components of this superposition
occupy different regions of space, the situation is completely different.\
The current variables $\mathbf{r}_{1},..,\mathbf{r}_{N}$ cannot sample
simultaneously all regions where $\left\langle D(\mathbf{r})\right\rangle $
is non-zero (if they did, they would sit at a point where the wave function $\Phi $ vanishes).\ The reason has already been discussed in \S~\ref{single-result}: because of the cohesive forces inside the object that enters a QSMDS (the
pointer of a measurement apparatus for instance), $\Phi $ vanishes unless
all $\mathbf{r}_{1},..,\mathbf{r}_{N}$ simultaneously belong to a single region of space
corresponding to one single component. By contrast, the sampling by $\left\langle D(\mathbf{r}%
)\right\rangle $ includes all regions.\ In this case,
the term in $\varepsilon $ of (\ref{GBC-31}) becomes large, so that the local
variation of the quantum density fluid $\partial \rho _{N}/\partial t$
significantly differs from the local variation of the density of Bohmian
variables $\bm\nabla _{N}\cdot \mathbf{J}_{N}$.\ The result is that, during a short transient
time, a large fraction of the quantum probability fluid is transferred
towards the region of lowest potential energy $\mathcal{V}_{G}(\mathbf{r})$
\ (largest value of its absolute value).\ In each realization of the
experiment, the Bohmian positions are insensitive to this transfer of probability density: they smoothly continue along their continuous
trajectories. In other words, it is now the Bohmian position that pilot the wave function.

As soon as the QSMDS\ is resolved by this process, one returns to the
previous regime where the Schr\"{o}dinger dynamics is an excellent
approximation.\ The usual relaxation process \cite%
{Valentini-Westman-2004,Towler-Russel-Valentini} takes place, and very soon
quantum equilibrium is restored.

\section{Conclusion\label{conclusion}}

In standard quantum mechanics, solutions of the Schr\"{o}dinger
equation become physically meaningless, when they have been left to
propagate \textquotedblleft too far\textquotedblright\ and involve a
coherent superposition of states that are macroscopically different and never observed.\ In standard theory, this problem is considered irrelevant: one postulates that the proper use of the equation
is to compute the evolution of quantum states only as long as they do not
involve QSMDS and macroscopic entanglement with the environment (for instance
during a quantum measurement); one merely ignores any further evolution of these
states when this condition is no longer met. 
\ The interest of modified Schr\"{o}%
dinger dynamics such as GRW, CSL, etc.\ is to propose a universal dynamics
that applies all the time, without any limit or selection of particular solutions of
the equation.

The form of the modified dynamics contained in GBC\ dynamics is
particularly simple. It relates the dynamical projection to a well-known field,
the gravity, and introduces no new dimensional constant. It quickly
resolves QSMDS if the quantum states of the superposition are
distinguishable in ordinary 3D space, corresponding to different spatial
densities of matter; by contrast, it would not resolve QSMDS occurring only
in momentum space, such as two macroscopic currents flowing in opposite
direction in a superconducting ring. This is because they correspond to the same distribution
of densities is space, which makes them insensitive to gravitational collapse.

Mathematically, what
determines the importance of the effects of a gravitational quantum collapse
is the quantum correlations of the local density with itself, or with other
observables.\ In ordinary space, and then in configuration space, we have
studied two situations: spontaneous appearance of quantum fluctuations of
density, or fluctuations induced by quantum measurement and described by a
transient QSMDS. The latter situation is simpler, since the
crossover region between the standard quantum regime and the fast localization
regime is crossed very suddenly, while
the entanglement quickly progresses towards larger and larger scales.
The standard von Neumann
projection \ postulate then appears as an extremely good approximation, a sort of \textquotedblleft sudden approximation\textquotedblright. In
the absence of quantum measurement, there is not reason why, in general,
such a rapid growth of the scale of entanglement should occur.\ The
localization process is then more complex, since it depends on the details
of various quantum correlation functions. In the frame of simple
approximations, we have seen that the gravitational localization process
leads to modified expressions of the gravity, either in a scalar form with a
change of the Newton constant $G$, or in a tensor form where the direction
of the Newton force is modified.

Obviously, the GBC model, as discussed in the present article, remains rather naive: it is
not expressed in terms of a field theory, and not even relativistic (it
relies on the standard dBB theory, which assumes a preferred reference
frame).\  As elementary as it is,  the model nevertheless illustrates how the enrichment of the
quantum dynamics, by adding a single point in the configuration space to the
$N$-body wave function, may lead to a completely different dynamics of
QSMDS.\ One may actually be surprised by the fact that such an elementary model should work so well, leading to an evolution of the state vector that is so similar to that of standard theory for microscopic systems, while also naturally including the von Neumann projection for QSMDS.
Moreover, since the GBC localization process is a continuous collective effect, the spontaneous heating of macroscopic systems is expected to be many orders of magnitude lower than that predicted by stochastic localization theories such as GRW of CSL. The GBC equations are particularly  simple, and do not require the introduction of stochastic processes or of the definition of a probability rule to specify them. Finally, if for instance one arbitrarily chooses the  fine structure constant for the value of $\varepsilon$, the dynamics requires the addition of no new physical constant into the equations.

From a historical point of view, one may notice some similarity with the ideas developed by
de Broglie when he worked on a double solution theory \cite%
{de-Broglie-1927, de-Broglie-1956}, which also involved a singular solution
localized at a precise point of configuration space. In GBC the singular
solution is replaced by a point $P$,
and the two
variables $\Psi $ and $P$ do not obey the same equation. Otherwise they
constantly guide each other, the reaction of $P$ on $\Psi $ being
significant only during the resolutions of QSMDS. Another analogy can be found with the \textquotedblleft shifty split\textquotedblright\ \cite{Mermin-shifty-split} contained in the Copenhagen interpretation, i.e. the moveable split (or cut)  between the quantum world of $S$  and the macroscopic world of measurement apparatuses $M$. We have seen that, because of the very rapid variation of the collapse time constant during the interaction between $S$ and $M$, the exact value of the constant $\varepsilon$ introduced in the localization term (\ref{grav-6}) is not so important: changing $\varepsilon$ shifts the time at which the sudden collapse occurs, but the change is only very small, so that a large range of values provides physically acceptable results. In other words, we can arbitrarily move the split, or cut, between the two worlds without changing much the predictions of the model.

Conceptually, the GBC\ model remains relatively neutral about interpretations and
ontology.\ Within the dBB theory, the traditional view is  that the
Bohmian positions provide the \textquotedblleft beables\textquotedblright\
\cite{Bell-local-beables, Bell-livre} of the theory.\ The state vector is then seen as a
mathematical object, similar to a Lagrangian, which pilots the motion of the
real positions.\ But an opposite point of view is possible witht
the GBC\ model.\ One can group all Bohmian positions into a single position
in configuration space and consider this position just as a mathematical
tool to enrich the quantum dynamics of spinless particles.\ Since the N-particle wave function constantly
follows the observations, and since every particle keeps a conditional wave function at any time in ordinary 3D space, one can consider that it directly represents the
physical reality of every particle, as initially envisaged by Schr\"{o}dinger when he
introduced the dynamical equation of wave mechanics.

\end{document}